\documentclass{article}
\usepackage[T1]{fontenc}
\usepackage[utf8]{inputenc}
\usepackage{ismir} \usepackage{lineno}
\usepackage{amsmath,amsfonts,cite,url}
\usepackage{multirow}
\usepackage{booktabs}
\usepackage{tabularx}
\usepackage{graphicx}
\usepackage{color}
\usepackage{cleveref}
\crefformat{figure}{#2Fig.~#1#3}
\crefmultiformat{figure}{Figs.~#2#1#3}{ and~#2#1#3}{, #2#1#3}{ and~#2#1#3}
\crefformat{equation}{#2Eq.~#1#3}
\crefmultiformat{equation}{Eqs.~#2#1#3}{ and~#2#1#3}{, #2#1#3}{ and~#2#1#3}

\title{Estimating Musical Surprisal from Audio \\ in Autoregressive Diffusion Model Noise Spaces}

\multauthor
  {Mathias Rose Bjare$^1$ \hspace{1cm} Stefan Lattner$^2$ \hspace{1cm} Gerhard Widmer$^{1,3}$}
  {$^1$ Institute of Computational Perception, Johannes Kepler University Linz, Austria\\ $^2$ Sony Computer Science Laboratories (CSL), Paris, France\\ $^3$ LIT AI Lab, Linz Institute of Technology, Austria\\ {\tt 
mathias.bjare@jku.at}
  }

\def\authorname{Mathias Rose Bjare, Stefan Lattner, and Gerhard Widmer}

\begin{document}

\maketitle

\begin{abstract}
Recently, the information content (IC) of predictions from a Generative Infinite-Vocabulary Transformer (GIVT) has been used to model musical expectancy and surprisal in audio. 
We investigate the effectiveness of such modelling using IC calculated with autoregressive diffusion models (ADMs).
We empirically show that IC estimates of models based on two different diffusion ordinary differential equations (ODEs) describe diverse data better, in terms of negative log-likelihood, than a GIVT. 
We evaluate diffusion model IC's effectiveness in capturing surprisal aspects by examining two tasks: (1) capturing monophonic pitch surprisal, and (2) detecting segment boundaries in multi-track audio. In both tasks, the diffusion models match or exceed the performance of a GIVT.
We hypothesize that the surprisal estimated at different diffusion process noise levels corresponds to the surprisal of music and audio features present at different audio granularities.
Testing our hypothesis, we find that, for appropriate noise levels, the studied musical surprisal tasks' results improve. 
Code is provided on 
\url{github.com/SonyCSLParis/audioic}.

\end{abstract}

\section{Introduction}
Surprisal, estimated via \textit{information content} (IC) or negative log-likelihood (NLL) of an autoregressive model, has been proposed as a proxy estimator for perceived musical surprise as experienced by human listeners \cite{meyer,idyom_conklin,idyom,musexp}. With suitable models, the IC of musical events correlates with human surprise perception and complexity, including tonal and rhythmic aspects  \cite{complic,BjareLW23}. Music analysis with IC enables quantitative information-theoretic hypotheses about music and music perception \cite{gold2019predictability}.
Furthermore, IC can serve as a conditioning signal for generative models \cite{wang2014guided,collins2016developing, bjare2024controlling}.
Recently, \cite{Bjare2024audioic} showed that surprisal modeling using IC calculated in the continuous audio latent space of \textit{Music2Latent} \cite{pasini2024music2latent} effectively models musical complexity, repetition reduction, and prediction of electroencephalogram (EEG) brain responses in human listeners. In
\cite{Bjare2024audioic}, a GIVT model \cite{Tshannen2023givt} is used that calculates IC using the likelihood of one-step predictions that are assumed to follow Gaussian mixture model (GMM) distributions with uncorrelated dimensions. However, this assumption may limit such models' effectiveness, given the nature of highly compressed continuous autoencoder representations, like Music2Latent.

Diffusion models have become powerful tools in generative AI, achieving state-of-the-art results in multiple domains, including music. A key advantage is that they do not rely on strong assumptions about how the data is distributed. Recent work \cite{SongD0S23} shows that by formulating diffusion processes as ODEs, a diffusion model can not only generate new samples but also estimate how likely (or “probable”) any given data point is under the model. Remarkably, this likelihood estimate can be computed at different stages of the diffusion process, which correspond to varying levels of “noise” or abstraction in the data.

In this paper, we study the ability of diffusion models to estimate musical surprisal.
We restrict our investigations to ADMs \cite{DBLP:conf/nips/LiTLDH24, pasini2024continuous}, since musical surprisal is causal in time. We experiment with diffusion models trained on multi-track audio following the popular EDM \cite{DBLP:conf/nips/KarrasAAL22} and the Rectified Flow \cite{DBLP:conf/iclr/LiuG023} processes, which differ in how they noise details.
We show that these models describe diverse music data better in terms of NLL than GIVT, consistent with audio fidelity results in generation tasks \cite{pasini2024continuous}.
We evaluate the estimated IC's effectiveness in modelling aspects of musical surprisal in audio on two tasks that have been studied in the monophonic symbolic domain:
capturing monophonic pitch surprisal, related to tonality understanding, and segment-boundary detection on multi-track audio, related to the information changes in music.
We show that surprisal estimated using diffusion models captures pitch surprisal better than the GIVT model. Furthermore, we demonstrate that peaks in the surprisal function align with segment boundaries; however, additional peaks are found.
Finally, we hypothesize that IC estimated at certain diffusion process noise levels can preserve the surprisal of higher-level audio features like pitch, while filtering out contributions to the IC of low-level features like timbre nuances.
We support our hypothesis by showing that, for appropriate noise levels, the results of the studied musical surprisal tasks improve.

\section{Related work}
\label{sec:related_work}
In the symbolic music domain, musical surprisal proxied by IC has most notably been studied with the variable order Markov-model IDyOM \cite{idyom_conklin}.
IDyOM modeling of human melodic expectation has been validated by numerous behavioral and neural studies  \cite{idyom, pitch_per, di2020cortical, hansen2014predictive, bianco2020pupil, moldwin2017statistical}.
The model is, however, limited to monophonic symbolic music stimuli. In \cite{bjare2024controlling}, the authors propose an IC-based technique for estimating surprisal in polyphonic symbolic music and show the IC to correlate with tonal and rhythmic complexity using solo piano performances.

In the audio domain, surprisal estimation typically relies on human-selected audio features. In \cite{abrams2022retrieving}, the IC of a D-REX \cite{Skerritt-DavisE18,skerritt2019model} model,  calculated using Bayesian inference, is related to the magnetoencephalography (MEG) brain response of human participants.
The Audio Oracle \cite{dubnov2007audio} analyzes surprisal using \textit{information rate} calculated from self-similarities of audio features and identifies high surprisal at segment boundaries.

Surprisal estimation using symbolic music or audio features faces two issues: the investigation is based on limited human-selected attributes, and a mismatch between what the computational model sees and what a human listener hears. Both cases potentially bias the investigation.

Therefore, most similar to our approach, \cite{Bjare2024audioic} estimates surprisal in an audio representation that preserves all features of the original audio. The authors estimate IC using the likelihood of a GIVT model \cite{Tshannen2023givt} and show that it can predict EEG responses to sung music. However, in contrast to ADMs, the GIVT model assumes that the next-step predictions follow a particular distribution, which may limit its predictive effectiveness.

Although unrelated to temporal surprisal, \cite{masclef2023deep} uses the KL-divergence of a diffusion model to approximate the likelihood of 5-second monophonic music clips. This is used to reproduce the inverted U-shape relation between the total IC of music clips and listener preference presented in  \cite{gold2019predictability}. The model does not rely on audio features; however, it ignores causality and memory aspects of surprisal.
In addition to IC, other measures of information have been proposed for the computational study of musical surprisal and expectation \cite{abdallah2009information,dubnov2021deep}. These, however, are impractical to calculate for continuous autoregressive models and have been limitedly validated perceptually in the literature.

\section{Method}

\subsection{Information Content Modelling}
Estimation of causal IC in a discrete (symbolic music) domain can be achieved effectively with GPT-style one-step prediction modelling. In this case, IC is calculated from the prediction target's log-likelihood according to an explicit (softmax) probability mass function with logits from a multi-layer perceptron (MLP) that takes as inputs a context state summarized by a causal Transformer model \cite{VaswaniSPUJGKP17}. As a result, the IC measures the likelihood of specific (musical) events.
In contrast, we aim to estimate the IC of \emph{continuous audio embeddings}, using the compressed representations of the \textit{Music2Latent} autoencoder \cite{pasini2024music2latent}.
In this continuous case, the probability mass cannot be modelled. Consequently, in \cite{Bjare2024audioic}, a GIVT model \cite{Tshannen2023givt} is used, which models the probability density of next-step predictions explicitly using a GMM with parameters from an MLP that takes as input a context state, summarized by a causal Transformer.
In this work, we do not require explicit density modeling. Instead, it suffices to obtain IC by \emph{log-likelihood point estimates} of the observations.
To that end, we calculate such point estimates using \textit{Autoregressive Diffusion Models (ADMs)} \cite{DBLP:conf/nips/LiTLDH24,pasini2024continuous}.
Similar to GPT-style causal transformers, ADMs summarize the context of past observations into a context state. However, instead of using an MLP to transform the context states into softmax logits or GMM parameters, ADMs use the context states to condition small diffusion model MLPs to \emph{generate} the next continuous state.

Estimating IC using a diffusion model requires the use of the Instantaneous Change of Variables formula \cite{DBLP:conf/nips/ChenRBD18}.
This formula, as detailed below, computes the log-likelihood of data points $\mathbf{z}_0\sim \pi_{0}$ (in our case, Music2Latent audio representations)
that can be flown to a known analytic distribution $\pi_{1}$ according to an ODE $\frac{d}{dt}\mathbf{z}(t)=f(\mathbf{z}(t), t)$.
Finding such ODEs is non-trivial, but it turns out that neural ODEs \cite{DBLP:conf/nips/ChenRBD18} derived from diffusion processes do exactly that: flow data samples to noise samples of known isotropic Gaussian distributions.

\subsection{Instantaneous Change of Variables}
\label{sec:changeofvariables}
For data points $z_0\sim \pi_0$ flowing in time $t$ according to $\frac{d}{dt}\mathbf{z}(t)=f(\mathbf{z}(t), t)$, \cite{DBLP:conf/nips/ChenRBD18} shows that the log-likelihood of points change according to another ODE: $
    \frac{d}{d t} \log p\left(\mathbf{z}\left(t\right)\right)= - \text{tr}\left(\frac{\partial f}{\partial\mathbf{z}}(\mathbf{z}(t), t)\right)
    $, given some regularity conditions.
Therefore, if $\mathbf{z}_0\sim\pi_0$ is flown to $\mathbf{z}_1 \sim \pi_1$ and $\pi_1$ is known, we can evaluate $\log\pi_0(\mathbf{z}_0)$ by the sum of $\log\pi_{1}(\mathbf{z}_1)$ and the log-likelihood flow change from $\pi_0$ to $\pi_1$. Pratically, the two ODEs are combined into a system of equations and solved numerical from $t_0$ to $t_{1}$ given the initial conditions $\mathbf{z}(t_0)=\mathbf{z}_{0}, \log \pi_{0}(\mathbf{z}_0) - \log \pi_{0}(\mathbf{z}(t_0))=0$:
\begin{equation}
\int_{t_0}^{t_1}
\underbrace{
\begin{bmatrix}
f(\mathbf{z}(t), t) \\
- \mathrm{tr} \left( \frac{\partial f}{\partial \mathbf{z}}(\mathbf{z}(t),t) \right)
\end{bmatrix}
}_{\text{dynamics}}
\, dt
=
\underbrace{
\begin{bmatrix}
\mathbf{z}_1 \\
\log \pi_{0}(\mathbf{z}_0) - \log \pi_{1}(\mathbf{z}_1)
\end{bmatrix}
}_{\text{solutions}}
.\label{eq:instchangeofvariables}
\end{equation}
We can now obtain $\log \pi_0(\mathbf{z}_{0})$ by adding $\log \pi_{1} (\mathbf{z}_1)$ to the solution of the 2nd ODE. The former can be easily evaluated using the solution of the 1st ODE and the known $\pi_1$.
Furthermore, \cite{DBLP:conf/iclr/GrathwohlCBSD19} shows that \cref{eq:instchangeofvariables} can be calculated efficiently with reverse-mode automatic differentiation using the Skilling-Hutchinson trace estimator \cite{skilling1989eigenvalues,hutchinson1989stochastic}, which involves using $n_r$ Monte Carlo runs with noise samples from a Rademacher distribution  \cite{hutchinson1989stochastic} to obtain an unbiased estimate of an expectation. For the approach to work, we therefore require finding ODEs that flow the data distribution $\pi_{0}$ to an analytic distribution $\pi_{1}$. In the following, we consider two diffusion model-based neural ODEs \cite{DBLP:conf/nips/ChenRBD18} learning such flows.

\subsection{Probability Flow ODEs}
\label{sec:probflowode}
In \cite{DBLP:conf/iclr/YangSohl21,DBLP:conf/nips/KarrasAAL22}, the authors define a diffusion noise (forward) process by a stochastic differential equation (SDE) that flow the data distribution $\pi_0$ to a (known) Gaussian distribution $\pi_1$ in time $t_0\rightarrow t_1$. The dynamics of the SDE on data points $\mathbf{z}(t_{0})$ flowing to $\mathbf{z}(t)$ can effectively be described by:
\begin{equation}
p_{t_{0},t}\left(\mathbf{z}(t) \vert \mathbf{z}(t_{0})\right) = \mathcal{N}\left(\mathbf{z}(t); \mathbf{z}(t_{0})s(t), s(t)^{2}\sigma(t)^{2} \mathbf{I}\right),
\label{eq:pertubationkernel}
\end{equation}
where $\sigma$ is a noise scale and $s$ a contraction chosen such that $p_{t_{0},t_{0}}\left(\mathbf{z}(t_0) \vert \mathbf{z}(t_{0})\right)=\pi_{0}$ and $p_{t_{0},t_{1}}\left(\mathbf{z}(t_1) \vert \mathbf{z}(t_{0})\right)=\pi_{1} \approx\mathcal{N}\left(\mathbf{z}(t_1); 0, \sigma_{\max}^{2} \mathbf{I}\right)$.
Remarkably, the SDEs can be translated to deterministic processes (probability flow ODEs) that equivalently flow  $\pi_0$ to $\pi_1$ given by:

\begin{equation}
\frac{d}{dt}\mathbf{z}(t) =s(t)^2 \dot{\sigma}(t){\sigma(t)} \nabla_{\mathbf{z}} \log p \left( \frac{\mathbf{z}(t)}{s(t)} ; \sigma(t) \right) -  \frac{\dot{s}(t)}{ s(t)} \mathbf{z}(t).
\label{eq:probflowode}
\end{equation}
These ODEs, thus, fulfil the requirements of \Cref{sec:changeofvariables}. 
\cref{eq:probflowode} can be turned into a neural ODE by learning $\nabla_{\mathbf{z}} \log p$ with a neural network (see \Cref{sec:architecture_train}) using score matching. In \Cref{sec:experiments}, we use the EDM initialization of \cref{eq:probflowode} from \cite{DBLP:conf/nips/KarrasAAL22}, where $s(t)=1$, $\sigma(t)=t$, and the process flows in time $t_0=0.002$ to $t_1=80$.
This model will be referred to as EDM in our experiments.

\subsection{Rectified Flow}
\label{sec:rff}
Rectified Flow (RFF) \cite{DBLP:conf/iclr/LiuG023} defines a process that flow the data distribution to a standard Gaussian distribution ($\pi_{1}=\mathcal{N}\left(\mathbf{z}(t_{1}); 0, \mathbf{I}\right)$) by following straight lines as much as possible. Formally, given the ODE: $\frac{d}{dt}\mathbf{z}(t) = v(\mathbf{z}(t), t)dt$
a, RFF between $\pi_{0}$ and $\pi_{1}$ is learned by the minimization:
\begin{equation}
    \min_{v} \int_{0}^{1} \mathbb{E} \left[ \left\| (\mathbf{z}_1- \mathbf{z}_0) - v(\mathbf{z}_t, t) \right\|^2 \right] dt,
    \label{eq:rectflowsmin}
\end{equation}
where $\mathbf{z}_0 \sim \pi_{0}, \mathbf{z}_1 \sim \pi_{1}$ and $\mathbf{z}_t = (1 - t)\mathbf{z}_0 + t\mathbf{z}_1$ for $t\in [0,1]$. $\mathbf{z}_t$ is therefore a point on the straight line connecting $\mathbf{z}_0, \mathbf{z}_1$. Substituting $v$ with a neural network (see \Cref{sec:architecture_train}) in the ODE, we get a neural ODE. The weights are learned using sample estimates of \cref{eq:rectflowsmin} as a loss.

\subsection{Likelihood Estimations in Noise Space Continua}
In addition to estimating likelihoods of the data distribution $\pi_{0}$ using the framework described above, we can also compute likelihoods at various noise levels along the noise/data continuum (that are traversed by varying $t$—either in the perturbation kernel of \cref{eq:pertubationkernel} for probability flow ODEs, or in $\mathbf{z}_t$ for RFF).
In both cases, rather than solving the ODE from $t_0$ to $t_1$, we solve it from the noise level $t$ down to $t_1$.

It is noted that both processes introduce noise gradually into the data. As a result, high-detail information is removed first, while lower-detail information is retained at lower noise levels, and then also lost as the noise increases. 
Therefore, in \Cref{sec:pitchsurprise} we hypothesize that the IC extracted at moderate noise levels captures the surprisal of certain lower-detail musical features—such as pitch—while filtering out the contributions of higher-detail features, like subtle timbral nuances.

Given a test example existing in $\pi_{0}$, there are three natural ways to obtain a "noised" version of the data at a given noise level $t$: (1) sampling from the noise process, (2) using the expected value of the noise process, or (3) solving the ODE from $t_0$ to $t$. We discard option (1) because it yields a stochastic estimate, and option (3) because we found that option (2)—using the expected value—produces better results in practice.

\section{Experiments and results}
\label{sec:experiments}
\subsection{Data}
\label{sec:data}
For training and evaluating our models, we use the following audio datasets and encode them into Music2Latent representations, using the public checkpoint of \cite{pasini2024music2latent}. 
For model training, we use a dataset consisting of 150,000 CC licenced full-length mixed-source MP3 files from Jamendo (JAM) \cite{jamendo}, which we split into 125k, 12.5k, and 12.5k examples for training, validating, and testing purposes, respectively.
For experiments involving monophonic singing voices, we use a private dataset for fine-tuning our models, comprising vocal stems from 20k songs.
For our experiments with monophonic pitch, we use a synthetic dataset (SYN) of 49 Irish folk tunes from \textit{The Session } dataset \cite{folkrnnsession}, synthesized at constant velocity with diverse SoundFont-based instruments according to the midi-programs: \textit{``Pad 1 (new age)''}, \textit{``Synth Voice''}, \textit{``Acoustic Guitar (nylon)''}, \textit{``Acoustic Grand Piano''} and \textit{``Trumpet''} for a total of 245 examples. Additionally, for each melody, the IC of IDyOM is computed (see \cite{BjareLW22} for details). Furthermore, we use the dataset of \cite{cantisani2023investigating} (VOC), consisting of 18 recorded vocal melodies paired with IDyOM IC estimates of the transcribed melodies. 
For our experiment on segment boundary detection, we use the Salami dataset (SAL) \cite{DBLP:conf/ismir/SmithBFRD11}, containing 1310 audio files, having segment annotations from one or two human annotators. The annotations are hierarchical and include the following levels: functional, uppercase, and lowercase, describing global structure with semantic segment labels, global structure, and local structures, respectively.
We use all datasets to evaluate model effectiveness using NLL. 

\subsection{Architecture and Training Details}
\label{sec:architecture_train}
For the diffusion models, we use a standard 12-layer causal Transformer backbone with Pre-Layer Normalization similar to \cite{radford2019language}, rotary positional embeddings \cite{Su2024Roformer}, and FlashAttention \cite{Dao2022Flashattention}, 
and for the diffusion MLP, we follow the architecture of \cite{pasini2024continuous}. For the GIVT model, we follow the architecture presented in \cite{Bjare2024audioic}.
All models are trained with a maximum sequence length of 4600, corresponding to approximately 7 minutes of audio and a batch size of 8 sequences, resulting in an effective batch size of up to $\sim$1 hour of music. We use Adam optimization for 270k steps with a learning rate of $3\cdot10^{-4}$ for the diffusion models and  $10^{-4}$ for GIVT, using a cosine schedule with a warmup of 1800 steps. For our experiments involving monophonic singing voices, we additionally fine-tune each model on the dataset mentioned in \Cref{sec:data} until convergence.

\subsection{ODE-based Likelihood Approximation Errors}

\begin{table}[t]
    \centering
                                                            
                                                                    \begin{tabular}{lcccccc}
        \toprule
        \multirow{3}{*}{S-MAE} & $\ n_r$ & 1 & 2 & 4 & 8 & 16 \\
        & EDM & .109 & .078 & .057 & .043 &  .033\\
        & RFF & .109 & .079 & .057 & .043 &  .033\\
        \cmidrule{1-7}
        \multirow{3}{*}{Q-MAE} & $\mathit{tol}$ & 1 & .1 &.01 & .001 & 1e-4 \\
        & EDM & .085 & .078 & .076 & .076 & .076 \\ 
        & RFF & .076 & .076 & .076 & .076 & .076 \\
        \midrule
                \multirow{3}{*}{Q-ME} & $\mathit{tol}$ & 1 & .1 & .01 & .001 & 1e-4\\
        & EDM & -.044 & -.018 & -.004 & .000 & .000 \\
        & RFF & .000 & -.001 & .000 & .000 & .000\\
        \bottomrule
    \end{tabular}
    \caption{Approximation errors of the likelihood estimation, indicated with the Skilling-Hutchinson (S) and quantization (Q) mean average error (MAE) and the quantization mean error (ME).  The error is reported with respect to references of $n_r=32$ runs and a tolerance of $tol=10^{-4}$ for the two error types, respectively. The results are normalized to the mean absolute NLL of the references.}
    \label{tab:hyp_comparison}
\end{table}

The likelihood estimation from ODE-based diffusion models is affected by two types of approximation errors: the discretization error of the ODEs and the Skilling-Hutchinson trace estimator (see \Cref{sec:changeofvariables}). In both cases, the approximation error can be controlled by trading off speed. We therefore perform initial experiments on 500 examples from our validation dataset to determine a suitable trade-off.
For the former, similarly to \cite{DBLP:conf/iclr/GrathwohlCBSD19,DBLP:conf/iclr/YangSohl21}, the error is controlled using the Runge-Kutta 5(4) \cite{dormand1980family} method.
For the latter, the unbiased approximation can be made arbitrarily small using enough Monte Carlo runs $n_{r}$. We use the scipy Runge-Kutta implementation \cite{2020SciPy-NMeth} with standard parameters except for setting the tolerances to $\mathit{atol} = \mathit{rtol}= \mathit{tol} = 10^{-3}$ and compute the mean absolute error (MAE) of the difference between NLL calculated with different $n_r$ and NLL of a reference calculated with a very large number of runs ($n_r=32$).

To relate the MAE to the scale of the NLL, we divide it by the reference's average absolute NLL and report the resulting measure as S-MAE in \Cref{tab:hyp_comparison}. For all $n_r$, we found that the average error is small for both models. Even when $n_r=1$, the error is $0.109$ of the average NLL. This demonstrates that it is possible to obtain a coarse estimate of a sample's NLL with minimal computational overhead compared to traditional diffusion model generation.
We identify $n_r=4$ as a good balance and fix it for further experiments.

For determining tolerance parameters $\mathit{tol}$, we similarly compare NLL calculated with different values of $\mathit{tol}$ to a reference of $\mathit{tol}=10^{-5}$ and report this as Q-MAE in \Cref{tab:hyp_comparison}. Furthermore, we investigate the bias by plotting the mean error (ME), normalized to the average NLL, and report it as Q-ME in \Cref{tab:hyp_comparison}. We find that for $\mathit{tol}\leq 0.1$ and $0.01$ for EDM and RFF, the absolute error does not improve compared to the reference. Comparing the bias of the error, we find that while RFF is unbiased, EDM has a negative bias for $tol>0.001$. The better performance of RFF is likely due to the straight flows imposed by the method, which allow the solver to take larger steps. Thus, we take $\mathit{tol}=0.001$, such that the relative MAE is $0.057$.

\subsection{Predictive Efficiency Comparision}
\label{sec:resmodelcomp}

\begin{table}[t]
    \centering
    \begin{tabular}{lccccc}
        \toprule
         & JAM & SAL & VOC & SYN\\
                GIVT & 0.925 & 1.053 & 1.182 & 0.981 \\
        EDM & 0.707 & \textbf{0.829} & \textbf{0.823} & \textbf{0.642} \\
        RFF & \textbf{0.699} & \textbf{0.829} & 0.831 & 0.656\\
                                        \bottomrule
    \end{tabular}
    \caption{Comparison of model NLLs in the Music2Latent on different datasets reported in bits/dimension.}
    \label{tab:nll_comparison}
\end{table}

Similar to previous work in density estimation models \cite{DBLP:conf/iclr/YangSohl21,DBLP:conf/icml/HoCSDA19,DBLP:conf/iclr/GrathwohlCBSD19}, we compare the model's predictive effectiveness (how well the models predict diverse audio data) using the average NLL reported in bits/dimension (mean negative $\log_2$-likelihood/dimension). Since all compared models estimate likelihoods in the fixed coordinate system of the Music2Latent codec, we can compare the NLL in that space. We emphasize that our reported results are, therefore, not directly comparable with those described in \cite{Bjare2024audioic} as it uses a different version of Music2Latent.

We find that \textit{Music2Latent} encodes silence into a small region of its latent space, causing the model to assign extremely low IC values to silent frames (since IC is unbounded from below for densities). Consequently, these low values downweight the average NLL calculations without improving the models' predictive capabilities. To address this, we remove leading and trailing silence from the audio before computing NLL.
Similarly, for each dataset, we discard the IC values at time steps that fall within the 1\% most extreme IC values (across any model). We present the models' average NLL in \Cref{tab:nll_comparison}.

We find that the diffusion models have much lower NLL than the GIVT model, and as such, model the one-step prediction densities more accurately. This is consistent with the findings of \cite{pasini2024continuous} for audio fidelity in a generative task. Comparing the NLL values of the EDM and RFF models reveals no clear winner. Interestingly, we find that the NLL of diffusion models on the SYN monophonic dataset, which is dissimilar to the training distribution JAM, is lowest. Using an information-theoretic interpretation, the low NLL indicates that the SYN dataset is less surprising regarding timbre and melody, and therefore has lower IC.

\subsection{Pitch Surprisal in Noise Space Continua}
\label{sec:pitchsurprise}
\begin{figure*}
    \centering
    \includegraphics[width=1.0\linewidth]{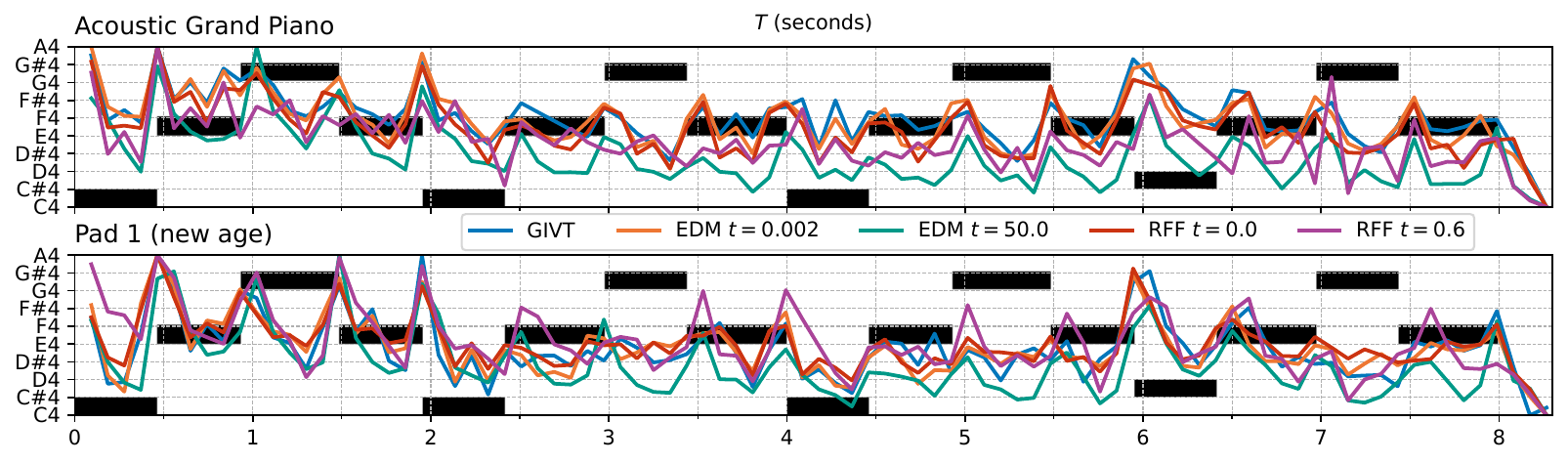}
    \caption{Piano roll of simple melody and IC of EDM and RFF models calculated from SoundFont audio synthesized at constant note-velocity with ``Acoustic Grand Piano'' (top) and ``Pad 1 (new age)'' (bottom) instruments. The IC is shown at noise levels $t=0.002,0.0$ (corresponding to fully unnoised data), and $t=50,0.6$ (corresponding to mid-process values) for EDM, RFF, respectively. The IC is affinely transformed to a min/max-value of $0/1$ for visualization.}
    \label{fig:entersurprise}
\end{figure*}

In \cite{pasini2024music2latent}, it is noted that a small MLP can predict pitches from the Music2Latent representations with high accuracy.
Thus, it is reasonable to hypothesize that pitch is embedded in coarse structures in the Music2Latent representations. We, therefore, investigate to what extent our IC can explain pitch surprisal and whether IC estimates at different noise levels describe pitch surprisal to a greater extent.

\subsubsection{IC Surprisal Qualitative}
\label{sec:qual}
In \cref{fig:entersurprise}, we plot a simple melody along with IC estimates on SoundFont audio synthesized with two different instruments.
The melody is composed of a $C$ major arpeggio repeated four times. On the fourth occurrence at $T=6$, the pattern is modified with the out-of-tonality pitch $C^{\#}$.
When predicting $E$ at $T=2.5s$ and forward, there is a partial match between the current context and the context seen $2s$ earlier.
That is, $E$ and the following notes are expected at time $T=2.5s$ and forward, which is reflected in the lower IC across all models in the investigated noise levels, compared to the ICs at times $T-2s$.
At $T=6s$, the surprising out-of-tonality note prompts a big peak in IC estimates across all models and noise levels. Finally, at time $T=8s$, the melody is surprisingly abruptly terminated, reflected in a smaller final peak in IC estimates across all models.
Comparing the IC estimates for the different timbre, we find that the mid-process IC estimates at $t=50.0, 0.6$, are more similar across timbre 
than the estimates of fully-denoised data at $t=0.002,0.0$ for EDM and RFF, respectively. This is evident, for example, by the lesser variation in IC between note onsets within $T=[0.5,1.0]$, which is especially visible for the EDM model. 
This suggests that the IC estimates at moderate noise levels capture pitch surprisal rather than timbre variations more effectively than estimates from fully unnoised data, which we investigate in the following.

\subsubsection{IC Surprisal Quantitative}
\begin{table}[t]
    \centering
    \begin{tabular}{lcccccc}
        \toprule
        \multirow{2}{*}{GIVT} & SYN & \multicolumn{4}{c}{-.031} \\
        & VOC & \multicolumn{4}{c}{.147} \\
        \cmidrule{1-7}
        \multirow{3}{*}{EDM} & $t$ & 2e-3 & 10.0 & 20.0 & 50.0 & 60.0 \\
                & SYN & .137 & .048 & .190 & \textbf{.264} & .255 \\
        & VOC & .135 & \textbf{.213} & .206 & .138 & .106 \\
        \cmidrule{1-7}
                                        \multirow{3}{*}{RFF} & $t$ & 0.0 & 0.1 & 0.5 & 0.6 & 0.7 \\
        & SYN & .134 & .189 & .216 & \textbf{.218} & .189 \\
        
        & VOC & \textbf{.137} & .133 & .069 & .048 & .030 \\
        \bottomrule
    \end{tabular}
    \caption{Spearman's correlation between IC calculated on (symbolic) melodies using IDyOM and IC calculated at different noise levels with EDM and RFF.}
    \label{tab:idyom}
\end{table}
To quantitatively validate whether the IC computed in audio can explain pitch surprisal, we would require a ground truth, which does not exist. As a proxy, we use the surprisal (IC) values predicted by the perceptually validated pitch expectancy model IDyOM (see \Cref{sec:related_work}), which operates in the symbolic domain.
We conduct our experiment with the SYN and VOC datasets. We extract the IDyOM IC of each note pitch in the symbolic datasets, and pair these with IC values calculated from our models on the audio datasets.
We align the latter to the notes by identifying the two Music2Latent frames that contain the note onset and calculating their average IC. 
Since we expect monotonic, rather than linear, relations in the IC pairs,
we compare the paired estimates using Spearman's rank correlation and report the results in \Cref{tab:idyom}. 
We find all correlations to be significant at a $5\%$-significance level and, except for GIVT, positive.
Comparing the GIVT to the diffusion models estimating IC of unnoised data ($t=0.002, 0.0$ for EDM and RFF, respectively), we find that the correlations are higher for SYN and similar for the VOC dataset.
In all cases, except for the RFF-VOC, we find the highest correlations using estimations with $t\neq t_0$, i.e., when IC is estimated using the noised data. For SYN, we see the highest correlations for high noise scale values $t=50.0, 0.6$ (compared to the fully noised noise values $t=80.0, 1.0$) for EDM and RFF, respectively. In particular, we find that EDM at noise level $t=50$ is overall mostly correlated with the IC of IDyOM, which supports the findings in \Cref{sec:qual} for similar data, where this value shows the smoothest surprisal curves, with clearly defined peaks around the note onsets.
For the VOC dataset, the highest correlations occur at lower noise levels, and overall, the correlations are less pronounced than in the SYN dataset.
This may be because singers are less precise when changing pitch, often using portamento to glide into a note.
As a result, peaks in IC during note changes may not be driven solely by pitch shifts, but also by other cues, such as emphasis at the onset (e.g., volume changes), vowel transitions, or the presence of plosives.
Masking such potentially subtle characteristics with noise may explain the observed correlation reduction.

\subsubsection{Noise Space Continua Timbre Invariance}
As shown above for SYN, the IC is more correlated with pitch surprisal at intermediate noise levels, where the fine details in the audio embeddings have been removed. We investigate to what extent this can be explained by a higher invariance to timbre irrelevant to pitch surprisal.
We, therefore, investigate if IC estimated on music that contains the same note content but with different timbre is more similar at the noise levels studied above. Specifically, we use SYN and investigate Spearman's correlation between IC of all combinations of pairs of synthetics sharing the same note material (but having different timbre), and report the results in \Cref{tab:selfsim}. The results are all significant at a $5\%$ significance level. For unnoised data, the correlations are similar for the diffusion models and GIVT. 
However, for EDM especially, the correlations increase for noised data.
We find high correlations for noise level values $t=50, 0.6$ for EDM and RFF, respectively, which have the highest correlation with IDyOM (See \cref{tab:idyom}), supporting the notion that these noise levels are more invariant to timbre.

\begin{table}[t]
    \centering
    \begin{tabular}{lccccc}
        \toprule
        GIVT & \multicolumn{5}{c}{.380}\\
        \midrule
         $t$ & 2e-3 & 10.0 & 20.0 & 50.0 & 60.0 \\
         EDM & 0.385 & 0.517 & \textbf{.525} & 0.522 & .518 \\
        \midrule
         $t$ & 0.0 & 0.1 & 0.5 & 0.6 & 0.7   \\
         RFF & 0.391 & 0.307 & .385 & \textbf{.429} & .402 \\
                                                                        \bottomrule
    \end{tabular}
    \caption{Spearman's correlation between IC estimations sharing the same note material, but with different timbre.}
    \label{tab:selfsim}
\end{table}

\subsection{IC for Unsupervised Segment Boundary Detection}
In the symbolic domain, IC has been used as a novelty function for segment boundary detection \cite{DBLP:series/sci/PearceMW10,DBLP:conf/mcm2/LattnerGAC15,DBLP:conf/ijcai/LattnerCG15}. Therefore, we investigate whether big changes in IC extracted from audio also coincide with segment boundaries. We conduct an experiment where we predict Salami lowercase segment boundaries using the most significant peaks extracted from an IC novelty function.
The novelty function is constructed by smoothing our IC curves with a Gaussian kernel with standard deviation $\sigma=5$, and differencing the smoothed series. Using the off-the-shelf Röder peak picking algorithm \cite{DBLP:journals/jossw/MullerZ21} with standard parameters, we report, in \Cref{tab:segmentboundary}, precision, recall, and $F_1$-score on predictions that are accurate within $\pm0.5$ seconds of the annotations. 
Generally, precision values are substantially lower than recall, implying that the IC novelty curves tend to have extra peaks not attributed to segment boundaries. 
For the GIVT, and the IC estimated with diffusion models on unnoised data, we find the $F_1$ scores to be similar.
For RFF, and to a lesser extent EDM, precision and recall increase with the noise level. This shows that the IC estimated at a coarser level aligns better with the segment boundaries.  
\begin{table}[t]
    \centering
    \begin{tabular}{lccccc}
        \toprule
        \multirow{3}{*}{GIVT} & prec & \multicolumn{4}{c}{.158} \\
         & rec & \multicolumn{4}{c}{.309}\\
         & $F_1$ & \multicolumn{4}{c}{.209}\\
         \midrule
         \multirow{4}{*}{EDM} & $t$ & 2e-3 & 17.6 & 40.0 & 60.0  \\
         & prec & .159 & .162 & .169 & \textbf{.178} \\
         & rec & .286 & .311 & .324 & \textbf{.345} \\
         & $F_1$ & .204 & .213 & .222 & \textbf{.235} \\
        \midrule
        \multirow{4}{*}{RFF} & $t$ & 0.0 & 0.25 & 0.50 & 0.70 \\
         & prec & .159 & .163 & .179 & \textbf{.198}  \\
         & rec & .287 & .342 & .380 & \textbf{.416} \\
         & $F_1$ & .205 & .221 & .243 & \textbf{.268} \\
        \bottomrule
    \end{tabular}
    \caption{Precision, recall and $F_1$ scores of Salami lowercase $\pm$0.5 seconds boundary detection.}
    \label{tab:segmentboundary}
\end{table}

\section{Conclusion and Discussion}
We investigated ADMs' ability to estimate musical surprisal and found that EDM and RFF diffusion models more effectively describe music data than a GITV in terms of NLL.
We evaluated the diffusion models IC's effectiveness in capturing monophonic pitch surprisal and found that these capture pitch surprisal better than a GIVT. Furthermore, we found that IC estimates of noised data increase correlation with pitch surprisal, and showed that this coincides with these estimates being more invariant to timbre.
Furthermore, we showed that peaks in a novelty function derived from IC coincide with Salami lowercase segment boundaries; however, the function has additional peaks. Finally, using the IC estimated in noise space improves the segment boundary predictions regarding precision and recall. 
As such, diffusion models surpass GIVT models in surprisal estimation and offer additional estimates that can capture aspects important to musical surprisal.

Similarly to \cite{Bjare2024audioic}, we estimate surprisal with IC in Music2Latent representations, so their findings on musical complexity, repetition reduction, and EEG prediction are likely to extend to diffusion-based IC. This should be validated in future work and extended with other perceptual validating experiments on diverse music.
Furthermore, our investigation of noise levels relevant to pitch surprisal could be extended to consider other perceptual features and their entanglement in different data representations. 
For instance, the IC calculated at suitable (high) noise levels in mel-spectrograms or constant-Q transformed representations
may give estimations of surprisal that correlate more with pitch surprisal.  
Also, the exploratory investigation of optimal noise levels could be  
automated using a methodology similar to \cite{daras2025ambient}, by monitoring performance degradations of a classifier/regressor model trained to predict the feature using variably noised inputs. Finally, our pitch surprisal analysis measured IC of coarse-grained structures, but our framework also allows studying surprising changes in fine-grained structures. This might, for instance, be relevant for analyzing timbre changes or singing techniques.

\clearpage

\section{Acknowledgments}
The work leading to these results was conducted in a collaboration between JKU and Sony Computer Science Laboratories Paris under a research agreement. The first and third author also acknowledge support by the European Research Council (ERC) under the European Union’s Horizon 2020 research and innovation programme, grant agreement 101019375 (\textit{“Whither Music?”}).
\bibliography{ISMIRtemplate}

\begin{thebibliography}{10}
\providecommand{\url}[1]{#1}
\csname url@samestyle\endcsname
\providecommand{\newblock}{\relax}
\providecommand{\bibinfo}[2]{#2}
\providecommand{\BIBentrySTDinterwordspacing}{\spaceskip=0pt\relax}
\providecommand{\BIBentryALTinterwordstretchfactor}{4}
\providecommand{\BIBentryALTinterwordspacing}{\spaceskip=\fontdimen2\font plus
\BIBentryALTinterwordstretchfactor\fontdimen3\font minus \fontdimen4\font\relax}
\providecommand{\BIBforeignlanguage}[2]{{%
\expandafter\ifx\csname l@#1\endcsname\relax
\typeout{** WARNING: IEEEtran.bst: No hyphenation pattern has been}%
\typeout{** loaded for the language `#1'. Using the pattern for}%
\typeout{** the default language instead.}%
\else
\language=\csname l@#1\endcsname
\fi
#2}}
\providecommand{\BIBdecl}{\relax}
\BIBdecl

\bibitem{meyer}
L.~B. Meyer, ``Meaning in music and information theory,'' \emph{The Journal of Aesthetics and Art Criticism}, vol.~15, no.~4, pp. 412--424, 1957.

\bibitem{idyom_conklin}
D.~Conklin and I.~H. Witten, ``Multiple viewpoint systems for music prediction,'' \emph{Journal of New Music Research}, vol.~24, no.~1, pp. 51--73, 1995.

\bibitem{idyom}
M.~Pearce, ``The construction and evaluation of statistical models of melodic structure in music perception and composition,'' Ph.D. dissertation, Department of Computing, City University, London, UK, 2005.

\bibitem{musexp}
M.~T. Pearce and G.~A. Wiggins, ``Auditory expectation: The information dynamics of music perception and cognition,'' \emph{Top. Cogn. Sci.}, vol.~4, no.~4, pp. 625--652, 2012.

\bibitem{complic}
S.~A. Sauv{\'e} and M.~T. Pearce, ``Information-theoretic modeling of perceived musical complexity,'' \emph{Music Perception: An Interdisciplinary Journal}, vol.~37, no.~2, pp. 165--178, 2019.

\bibitem{BjareLW23}
M.~R. Bjare, S.~Lattner, and G.~Widmer, ``Exploring sampling techniques for generating melodies with a transformer language model,'' in \emph{{ISMIR}}, 2023, pp. 810--816.

\bibitem{gold2019predictability}
B.~P. Gold, M.~T. Pearce, E.~Mas-Herrero, A.~Dagher, and R.~J. Zatorre, ``Predictability and uncertainty in the pleasure of music: a reward for learning?'' \emph{Journal of Neuroscience}, vol.~39, no.~47, pp. 9397--9409, 2019.

\bibitem{wang2014guided}
C.-i. Wang and S.~Dubnov, ``Guided music synthesis with variable markov oracle,'' in \emph{{AAAI}}, vol.~10, no.~5, 2014, pp. 55--62.

\bibitem{collins2016developing}
T.~Collins, R.~Laney, A.~Willis, and P.~H. Garthwaite, ``Developing and evaluating computational models of musical style,'' \emph{AI EDAM}, vol.~30, no.~1, pp. 16--43, 2016.

\bibitem{bjare2024controlling}
M.~R. Bjare, S.~Lattner, and G.~Widmer, ``Controlling surprisal in music generation via information content curve matching,'' in \emph{{ISMIR}}, 2024.

\bibitem{Bjare2024audioic}
M.~R. Bjare, G.~Cantisani, S.~Lattner, and G.~Widmer, ``Estimating musical surprisal in audio,'' in \emph{{ICASSP}}, 2025.

\bibitem{pasini2024music2latent}
M.~Pasini, S.~Lattner, and G.~Fazekas, ``Music2latent: Consistency autoencoders for latent audio compression,'' in \emph{{ISMIR}}, 2024.

\bibitem{Tshannen2023givt}
M.~Tschannen, C.~Eastwood, and F.~Mentzer, ``{GIVT:} generative infinite-vocabulary transformers,'' \emph{CoRR}, vol. abs/2312.02116, 2023.

\bibitem{SongD0S23}
Y.~Song, P.~Dhariwal, M.~Chen, and I.~Sutskever, ``Consistency models,'' in \emph{{ICML}}, vol. 202, 2023, pp. 32\,211--32\,252.

\bibitem{DBLP:conf/nips/LiTLDH24}
T.~Li, Y.~Tian, H.~Li, M.~Deng, and K.~He, ``Autoregressive image generation without vector quantization,'' in \emph{NeurIPS}, 2024.

\bibitem{pasini2024continuous}
M.~Pasini, J.~Nistal, S.~Lattner, and G.~Fazekas, ``Continuous autoregressive models with noise augmentation avoid error accumulation,'' in \emph{Audio Imagination: NeurIPS 2024 Workshop AI-Driven Speech, Music, and Sound Generation}, 2024.

\bibitem{DBLP:conf/nips/KarrasAAL22}
T.~Karras, M.~Aittala, T.~Aila, and S.~Laine, ``Elucidating the design space of diffusion-based generative models,'' in \emph{NeurIPS}, 2022.

\bibitem{DBLP:conf/iclr/LiuG023}
X.~Liu, C.~Gong, and Q.~Liu, ``Flow straight and fast: Learning to generate and transfer data with rectified flow,'' in \emph{{ICLR}}, 2023.

\bibitem{pitch_per}
M.~T. Pearce, M.~H. Ruiz, S.~Kapasi, G.~A. Wiggins, and J.~Bhattacharya, ``Unsupervised statistical learning underpins computational, behavioural, and neural manifestations of musical expectation,'' \emph{NeuroImage}, vol.~50, no.~1, pp. 302--313, 2010.

\bibitem{di2020cortical}
G.~M. Di~Liberto, C.~Pelofi, R.~Bianco, P.~Patel, A.~D. Mehta, J.~L. Herrero, A.~De~Cheveign{\'e}, S.~Shamma, and N.~Mesgarani, ``Cortical encoding of melodic expectations in human temporal cortex,'' \emph{Elife}, vol.~9, p. e51784, 2020.

\bibitem{hansen2014predictive}
N.~C. Hansen and M.~T. Pearce, ``Predictive uncertainty in auditory sequence processing,'' \emph{Frontiers in psychology}, vol.~5, p. 1052, 2014.

\bibitem{bianco2020pupil}
R.~Bianco, L.~E. Ptasczynski, and D.~Omigie, ``Pupil responses to pitch deviants reflect predictability of melodic sequences,'' \emph{Brain and Cognition}, vol. 138, p. 103621, 2020.

\bibitem{moldwin2017statistical}
T.~Moldwin, O.~Schwartz, and E.~S. Sussman, ``Statistical learning of melodic patterns influences the brain's response to wrong notes,'' \emph{Journal of cognitive neuroscience}, vol.~29, no.~12, pp. 2114--2122, 2017.

\bibitem{abrams2022retrieving}
E.~Abrams, E.~M. Vidal, C.~Pelofi, and P.~Ripoll{\'e}s, ``Retrieving musical information from neural data: how cognitive features enrich acoustic ones.'' in \emph{{ISMIR}}, 2022, pp. 160--168.

\bibitem{Skerritt-DavisE18}
B.~Skerritt{-}Davis and M.~Elhilali, ``Detecting change in stochastic sound sequences,'' \emph{PLoS Comput. Biol.}, vol.~14, no.~5, 2018.

\bibitem{skerritt2019model}
B.~Skerritt-Davis and M.~Elhilali, ``A model for statistical regularity extraction from dynamic sounds,'' \emph{Acta Acustica united with Acustica}, vol. 105, no.~1, pp. 1--4, 2019.

\bibitem{dubnov2007audio}
S.~Dubnov, G.~Assayag, and A.~Cont, ``Audio oracle: A new algorithm for fast learning of audio structures,'' in \emph{{ICMC}}, 2007, pp. 224--227.

\bibitem{masclef2023deep}
N.~L. Masclef and T.~A. Keller, ``Deep generative models of music expectation,'' \emph{NeurIPS ML for Audio Workshop 2023}, 2023.

\bibitem{abdallah2009information}
S.~Abdallah and M.~Plumbley, ``Information dynamics: patterns of expectation and surprise in the perception of music,'' \emph{Connection Science}, vol.~21, no. 2-3, pp. 89--117, 2009.

\bibitem{dubnov2021deep}
S.~Dubnov, ``Deep music information dynamics,'' \emph{arXiv preprint arXiv:2102.01133}, 2021.

\bibitem{VaswaniSPUJGKP17}
A.~Vaswani, N.~Shazeer, N.~Parmar, J.~Uszkoreit, L.~Jones, A.~N. Gomez, L.~Kaiser, and I.~Polosukhin, ``Attention is all you need,'' in \emph{NeurIPS}, 2017, pp. 5998--6008.

\bibitem{DBLP:conf/nips/ChenRBD18}
T.~Q. Chen, Y.~Rubanova, J.~Bettencourt, and D.~Duvenaud, ``Neural ordinary differential equations,'' in \emph{NeurIPS}, 2018, pp. 6572--6583.

\bibitem{DBLP:conf/iclr/GrathwohlCBSD19}
W.~Grathwohl, R.~T.~Q. Chen, J.~Bettencourt, I.~Sutskever, and D.~Duvenaud, ``{FFJORD:} free-form continuous dynamics for scalable reversible generative models,'' in \emph{{ICLR}}, 2019.

\bibitem{skilling1989eigenvalues}
J.~Skilling, ``The eigenvalues of mega-dimensional matrices,'' \emph{Maximum Entropy and Bayesian Methods: Cambridge, England, 1988}, pp. 455--466, 1989.

\bibitem{hutchinson1989stochastic}
M.~F. Hutchinson, ``A stochastic estimator of the trace of the influence matrix for laplacian smoothing splines,'' \emph{Communications in Statistics-Simulation and Computation}, vol.~18, no.~3, pp. 1059--1076, 1989.

\bibitem{DBLP:conf/iclr/YangSohl21}
Y.~Song, J.~Sohl{-}Dickstein, D.~P. Kingma, A.~Kumar, S.~Ermon, and B.~Poole, ``Score-based generative modeling through stochastic differential equations,'' in \emph{{ICLR}}, 2021.

\bibitem{jamendo}
Jamendo, ``{Jamendo Music},'' \url{https://www.jamendo.com}.

\bibitem{folkrnnsession}
B.~L. Sturm, J.~F. Santos, O.~Ben-Tal, and I.~Korshunova, ``Music transcription modelling and composition using deep learning,'' in \emph{Proceedings of the Conference on Computer Simulation of Musical Creativity}, Huddersfield,{UK}, 2016.

\bibitem{BjareLW22}
M.~R. Bjare, S.~Lattner, and G.~Widmer, ``Differentiable short-term models for efficient online learning and prediction in monophonic music,'' \emph{Trans. Int. Soc. Music. Inf. Retr.}, vol.~5, no.~1, p. 190, 2022.

\bibitem{cantisani2023investigating}
G.~Cantisani, A.~Chalehchaleh, G.~Di~Liberto, and S.~Shamma, ``Investigating the cortical tracking of speech and music with sung speech,'' in \emph{{INTERSPEECH}}.\hskip 1em plus 0.5em minus 0.4em\relax {ISCA}, 2023, pp. 5157--5161.

\bibitem{DBLP:conf/ismir/SmithBFRD11}
J.~B.~L. Smith, J.~A. Burgoyne, I.~Fujinaga, D.~D. Roure, and J.~S. Downie, ``Design and creation of a large-scale database of structural annotations,'' in \emph{{ISMIR}}, 2011, pp. 555--560.

\bibitem{radford2019language}
A.~Radford, J.~Wu, R.~Child, D.~Luan, D.~Amodei, I.~Sutskever \emph{et~al.}, ``Language models are unsupervised multitask learners,'' \emph{OpenAI blog}, vol.~1, no.~8, p.~9, 2019.

\bibitem{Su2024Roformer}
J.~Su, M.~H.~M. Ahmed, Y.~Lu, S.~Pan, W.~Bo, and Y.~Liu, ``Roformer: Enhanced transformer with rotary position embedding,'' \emph{Neurocomputing}, vol. 568, p. 127063, 2024.

\bibitem{Dao2022Flashattention}
T.~Dao, D.~Y. Fu, S.~Ermon, A.~Rudra, and C.~R{\'{e}}, ``Flashattention: Fast and memory-efficient exact attention with io-awareness,'' in \emph{NeurIPS}, 2022.

\bibitem{dormand1980family}
J.~R. Dormand and P.~J. Prince, ``A family of embedded runge-kutta formulae,'' \emph{Journal of computational and applied mathematics}, vol.~6, no.~1, pp. 19--26, 1980.

\bibitem{2020SciPy-NMeth}
P.~Virtanen, R.~Gommers, T.~E. Oliphant, M.~Haberland, T.~Reddy, D.~Cournapeau, E.~Burovski, P.~Peterson, W.~Weckesser, J.~Bright, S.~J. {van der Walt}, M.~Brett, J.~Wilson, K.~J. Millman, N.~Mayorov, A.~R.~J. Nelson, E.~Jones, R.~Kern, E.~Larson, C.~J. Carey, {\.I}.~Polat, Y.~Feng, E.~W. Moore, J.~{VanderPlas}, D.~Laxalde, J.~Perktold, R.~Cimrman, I.~Henriksen, E.~A. Quintero, C.~R. Harris, A.~M. Archibald, A.~H. Ribeiro, F.~Pedregosa, P.~{van Mulbregt}, and {SciPy 1.0 Contributors}, ``{{SciPy} 1.0: Fundamental Algorithms for Scientific Computing in Python},'' \emph{Nature Methods}, vol.~17, pp. 261--272, 2020.

\bibitem{DBLP:conf/icml/HoCSDA19}
J.~Ho, X.~Chen, A.~Srinivas, Y.~Duan, and P.~Abbeel, ``Flow++: Improving flow-based generative models with variational dequantization and architecture design,'' in \emph{{ICML}}, vol.~97, 2019, pp. 2722--2730.

\bibitem{DBLP:series/sci/PearceMW10}
M.~T. Pearce, D.~M{\"{u}}llensiefen, and G.~A. Wiggins, ``Melodic grouping in music information retrieval: New methods and applications,'' in \emph{Advances in Music Information Retrieval}, 2010, vol. 274, pp. 364--388.

\bibitem{DBLP:conf/mcm2/LattnerGAC15}
S.~Lattner, M.~Grachten, K.~Agres, and C.~E.~C. Chac{\'{o}}n, ``Probabilistic segmentation of musical sequences using restricted boltzmann machines,'' in \emph{{MCM}}, vol. 9110, 2015, pp. 323--334.

\bibitem{DBLP:conf/ijcai/LattnerCG15}
S.~Lattner, C.~E.~C. Chac{\'{o}}n, and M.~Grachten, ``Pseudo-supervised training improves unsupervised melody segmentation,'' in \emph{{IJCAI}}.\hskip 1em plus 0.5em minus 0.4em\relax {AAAI} Press, 2015, pp. 2459--2465.

\bibitem{DBLP:journals/jossw/MullerZ21}
\BIBentryALTinterwordspacing
M.~M{\"{u}}ller and F.~Zalkow, ``libfmp: {A} python package for fundamentals of music processing,'' \emph{J. Open Source Softw.}, vol.~6, no.~63, p. 3326, 2021. [Online]. Available: \url{https://doi.org/10.21105/joss.03326}
\BIBentrySTDinterwordspacing

\bibitem{daras2025ambient}
G.~Daras, A.~Rodriguez-Munoz, A.~Klivans, A.~Torralba, and C.~Daskalakis, ``Ambient diffusion omni: Training good models with bad data,'' \emph{arXiv preprint arXiv:2506.10038}, 2025.

\end{thebibliography}

\end{document}